\documentclass[10pt,a4paper]{amsart}
\usepackage{amssymb}
     \def\p{\partial}
     
     \def\bi{\begin{itemize}}
     \def\ei{\end{itemize}}
     
     \def\be{\begin{equation}}
     \def\ee{\end{equation}}
     \def\ph{\phantom{x}}
     \def\h{\hfill}

     \def\h{\hat}
\newcommand{\Rnos}{{\rm I \mkern-2.5mu \nonscript\mkern-.5mu R}}

     \newcommand{\vb}[1]{\mbox{\boldmath{$#1$}}}
     \newcommand{\uvb}[1]{\mbox{\boldmath{$\hat{#1}$}}}
     \newcounter{qnum}
\theoremstyle{plain}
\newtheorem{theorem}{Theorem}
\newtheorem{corollary}{Corollary}

\newtheorem{lemma}{Lemma}

\newtheorem{definition}{Definition}

\theoremstyle{remark}

\begin{document}
\title[Magnetic Monopoles, Electric Neutrality]
{Magnetic Monopoles, Electric Neutrality\\
      and the Static Maxwell-Dirac Equations}

\author{Chris Radford}
\address{School of Mathematical and Computer Sciences\\
         University of New England\\
         Armidale NSW 2351\\
          Australia}
\email{chris@turing.une.edu.au}

\author{Hilary Booth}
\address{Pure and Applied Mathematics\\
         University of Adelaide\\
         Adelaide SA 5005
          Australia}
\email{hbooth@maths.adelaide.edu.au}
\flushleft
\date{\today}

\begin{abstract}
We study the full Maxwell-Dirac equations: Dirac field with minimally coupled electromagnetic field and Maxwell field with Dirac 
current as source. Our particular interest is the static case in which the Dirac current is purely time-like -- the ``electron" is at 
rest in some Lorentz frame. In this case we prove two theorems under rather general assumptions. Firstly, that if the system is also 
stationary (time independent in some gauge) then the system as a whole must have vanishing total charge, i.e. it must be electrically 
neutral. In fact, the theorem only requires that the system be {\em asymptotically} stationary and static. 
Secondly, we show, in the axially symmetric case, that if there are external Coulomb fields then these must necessarily be 
magnetically charged -- all Coulomb external sources are electrically charged magnetic monopoles.

\end{abstract}
\maketitle

\newpage
\flushleft
{\bf I. Introduction}

The Maxwell-Dirac equations are the classical field (or more ``traditionally", first quantised) equations for electronic matter. 
Historically, only the linearised equations (where the Dirac current is ignored as a source for the Maxwell equations) have been 
studied in detail -- for a comprehensive survey of the Dirac equation with various potentials see Thaller \cite{BT:dbook}. The lack of 
past interest in the full Maxwell-Dirac equations is partly due to the very difficult nonlinearities of the equations. More importantly
, the classical problem was swamped by the extraordinary success of QED.

The difficult nature of these nonlinear equations has meant that the existence theory has only recently been enunciated -- some 
highlights in this development might be Gross \cite{LG:cauch}, Chadam \cite{JC:cauch}, Georgiev \cite{VG:exs}, Esteban {\em et al} 
\cite{EGS:stexs}, and Bournaveas \cite{NB:exs}. This work culminated in a tour de force of nonlinear functional analysis, the global 
existence proof of Flato, 
Simon and Taflin \cite{FST:gexs}.

Our aim in studying the Maxwell-Dirac system is to look for possible non-linear behaviour which would not be apparent in perturbation 
expansions. The particular solutions found in \cite{CR:pap1} and \cite{HC:pap2} exhibit just this sort of behaviour -- localisation and 
charge screening. See also Das \cite{AD:md} and the recent work of Finster, Smoller and Yau \cite{FSY:mds}.

The static Maxwell-Dirac equations were first written down in \cite{CR:pap1}. In the present work we use this formulation to prove two 
theorems. Firstly, that the stationary, static Maxwell-Dirac system must have vanishing total charge, this is done in section IV. The 
second theorem proves that, in the axially symmetric case an external Coulomb field must have an associated magnetic charge -- 
external Coulomb fields must be electrically charged magnetic monopoles. This theorem is proved in section VI.

{\bf II. The Static Maxwell-Dirac Equations}

   In standard notation the Dirac-Maxwell equations are
    \begin{eqnarray*}
 &&  \gamma^\alpha (\partial_\alpha -i\,e\,A_\alpha )\psi +
im\psi =0\\
 && F_{\alpha \beta} = A_{\beta ,\alpha} - A_{\alpha , \beta}\\
     &&\partial ^\alpha F_{\alpha \beta} = -4\pi e\,j_\beta = -4\pi e\,\overline
{\psi}
\gamma_\beta \psi
    \end{eqnarray*}
In \cite{CR:pap1} the 2-spinor form of the Dirac equations was employed to solve for the electromagnetic potential, under the 
non-degeneracy condition $j^{\alpha}j_{\alpha}\neq 0$. Requiring $A^{\alpha}$ to be a real four-vector then gave  a set of partial 
differential equations in the Dirac field alone, the reality conditions.

For 2-spinors $u_A$ and $v^B$ we have
     \begin{eqnarray*}
&& \psi = \left(\begin{array}{c}
     u_A\\
     \overline{v}^{\dot{B}}
     \end{array}
     \right),  \ \ \mbox{with}\\
&&   u_Cv^C \neq 0 \ \ \mbox{(non-degeneracy)}. 
      \end{eqnarray*}
The electromagnetic potential,
\begin{eqnarray*}
 A^{A\dot{A}}&=&\frac{i}{e(u^cv_c)}\left\{
v^A\partial ^{B\dot{A}}
u_B+u^A\partial^{B\dot{A}}v_B+\frac{im}{\sqrt{2}}(u^A\overline{u}^{\dot{A}}+v^A
\overline{v}^{\dot{A}})\right\}.
\end{eqnarray*}
The reality conditions,
\begin{eqnarray*}
 &&\partial^{A\dot{A}}(u_A\overline{u}_{\dot{A}}) = -
\frac{im}{\sqrt{2}}(u^C v_C -\overline{u}^{\dot{C}}\overline{v}_{\dot{C}})\\
 && \partial^{A\dot{A}}(v_A\overline{v}_A)= \frac{im}{\sqrt{2}} (u^C v_
C -
\overline{u}^{\dot{C}}\overline{v}_{\dot{C}})\\
 &&u_A\partial^{A\dot{A}}\bar{v}_{\dot{A}} -
\overline{v}_{\dot{A}}\partial^{A\dot{A}} u_A=0
     \end{eqnarray*}
The Maxwell equations,
\begin{eqnarray*}
\partial ^{\alpha}F_{\alpha \beta}
=-4\pi e\,j_\beta =
-4\pi e\,\sigma_\beta^{A\dot{A}}(u_A\overline{u}_{\dot{A}}+v_A\overline{v}_{\dot
{A}}).
\end{eqnarray*}
These equations constitute the Maxwell-Dirac system.

We next impose the static condition.
\begin{definition} The Maxwell-Dirac system is said to be static if there exists a local Lorentz frame in which the Dirac current 
vector is purely time-like, i.e. $j^{\alpha}=j^0\delta^{\alpha}_{0}$.
\end{definition}
Imposing this condition one quickly finds that
     \[ v^A=e^{i\chi}
\sqrt{2}\sigma^{0 A\dot{A}}\overline{u}_{\dot{A}},\ \mbox{with $\chi$ a real
function.} \]
The current vector is now
     \[ j^\alpha = \sqrt{2}(u^0\overline{u}^{\dot{0}}+u^1
\overline{u}^{\dot{1}})\delta^\alpha_0 .\]

As noted in \cite{CR:pap1} the gauge is fixed by the choice,
     \begin{eqnarray}\label{f-gauge}
\nonumber  u^0  &=& X\,e^{\frac{i}{2}(\chi + \eta )}\\
           u^1&=& Y\,e^{\frac{i}{2}(\chi -\eta )},
     \end{eqnarray}
with $X$, $Y$, and $\eta$ real functions on $\Rnos^4$.

Defining the null vector $L$,
     \begin{eqnarray*}
      L & = & (\sigma^{\alpha}_{ A \dot{A}} u^A \bar{u}^{\dot{A}})
              =  (L^0,\frac{1}{\sqrt{2}}\vb{V}),\ \ \mbox{with}\ \ 
     L^0 = \frac{1}{\sqrt{2}} (X^2+Y^2) \ \ \mbox{and} \\
     \vb{V} &=& (2 XY\cos \eta, \ 2XY\sin \eta, X^2-Y^2),
     \end{eqnarray*}

our equations become,
\begin{eqnarray*}
&&\frac{\partial}{\partial t} (X^2+Y^2)=0\\
&&     \vb{\nabla .V} = -2m(X^2+Y^2)\sin \chi\\
&&     \frac{\partial \vb{V}}{\partial t} + (\vb{\nabla}\chi)\vb{\times V}=\vb{0}.
     \end{eqnarray*}
With electromagnetic potential
    \begin{eqnarray*}
   &&A^0 =\frac me\cos \chi +\frac{(X^2-
Y^2)}{2\,e(X^2+Y^2)}\frac{\partial\eta }{\partial
t}+\frac{(\vb{\nabla} \chi)\vb{. V}}{2e(X^2 + Y^2)}\\
     &&\vb{A} = \frac{1}{2e(X^2 + Y^2)} \left[ \frac{\partial
\chi}{\partial t} \vb{V} + (X^2 - Y^2) \vb{\nabla} \eta
-\vb{\nabla \times V} \right]\\
     &&{\rm where} \ \ \vb{A} = (A^1, A^2, A^3).
     \end{eqnarray*}
The full system is given by the above two sets of equations and the Maxwell
equations.

One further condition we want to impose is that of stationarity.
\begin{definition} The Maxwell-Dirac system will be called stationary if there exists a gauge in which $\psi  =e^{i\omega t}\phi$, 
with the bi-spinor $\phi$ independent of $t$. Such a gauge will be referred to as a stationary gauge.
\end{definition}
We now have the following simple lemma.
\begin{lemma} The static Maxwell-Dirac system is stationary if and only if, in the gauge given in \eqref{f-gauge}, 
\[\frac{\p \eta}{\p t}=0\ \ \mbox{and}\ \ \frac{\p X}{\p t}=0.\]
In the stationary case we also have,
\[ \frac{\p \chi}{\p t}=0\ \ \mbox{and}\ \ \frac{\p\vb{V}}{\p t}=\vb{0}.\]
\end{lemma}
\begin{proof} If the system is stationary there exists a gauge transformation such that
\[ u^A\to e^{i\xi}u^A=e^{i\omega t}\zeta^A,\ \ \mbox{with}\ \ \zeta^A\ \ \mbox{independent of $t$}.\]
Consequently,
\begin{eqnarray*}
Xe^{\frac{i}{2}(\chi +\eta)}&=&e^{i(\omega t-\xi)}\zeta^0\\
Ye^{\frac{i}{2}(\chi -\eta)}&=&e^{i(\omega t-\xi)}\zeta^1.
\end{eqnarray*}
So, $|X|=|\zeta^0|$ and $|Y|=|\zeta^1|$, both independent of $t$. We also have,
\begin{eqnarray*}
\frac{\p \chi}{\p t}+\frac{\p \eta}{\p t}&=&2(\omega-\frac{\p\xi}{\p t})\\
\frac{\p \chi}{\p t}-\frac{\p \eta}{\p t}&=&2(\omega-\frac{\p\xi}{\p t}).
\end{eqnarray*}
So that, $\frac{\p\eta}{\p t}=0$. The argument is easily reversed to get the converse statement.
\end{proof}

{\bf III. Isolated Systems} 

An isolated system is one for which all sources are contained in some ball $B_\rho$ ($\rho <\infty$) and for which the fields die off as $|x|=r\to\infty$.
 
In what follows we will be considering stationary Maxwell-Dirac systems. For such systems we would expect, in an appropriate stationary gauge, that 
$A^\alpha$ should be $O(1/r)$ as $r\to\infty$. We will also need to impose some decay conditions on $\psi$ as $r\to \infty$ in order to appropriately 
define an isolated Maxwell-Dirac system. The best language for the discussion of such decay conditions and other regularity issues is the language of 
weighted function spaces; specifically weighted classical and Sobolev spaces.

We will use the definitions of \cite{RB:mass}, other accounts of the theory may be 
found in \cite{CC:sob}, \cite{JC:sob} and \cite{HKM:sob}.
\begin{definition}
Weighted Sobolev spaces can be defined via the weighted Lebesgue spaces $L^p_\delta$, $1\leq p\leq\infty$ which are 
spaces of locally measurable functions for which the norms
\[ \|f\|_{p,\delta}=\left\{\begin{array}{l}
\left(\int_{\Rnos^n}|f|^p\sigma^{-p\delta -n}\, dx\right)^{\frac{1}{p}},\ \ p<\infty\\
\mbox{\em ess sup}_{\Rnos^n}\left( \sigma^{-\delta}|f|\right),\ \ p=\infty ,\end{array}\right.\]
are finite. We can replace $\Rnos^n$ with subsets $\Omega$ of $\Rnos^n$ in these definitions. The weight $\sigma$ 
is usually taken to be $\sigma =\sqrt{1+r^2}$ or $\sigma =r$ on subsets excluding $\{ 0\}$. The weighted Sobolev 
spaces are now defined as consisting of functions with weak derivatives up to order $k$ for which the following 
norm is finite
\[ \|f\|_{k,p,\delta} =\sum^k_{j=0}\|D^jf\|_{p,\delta -j}.\]
For $p<\infty$ we denote the weighted Sobolev space by $W^{k,p}_\delta$. For $p=\infty$ we denote the classical weighted function space by $C^k_\delta$.
\end{definition}

We will also require the Sobolev inequality, given here in the form presented in \cite{RB:mass}.

{\bf Sobolev Inequality}

{\em If $F\in W^{k,p}_\delta$ then
\[ \begin{array}{ll}
\mbox{(i)}&\|f\|_{\frac{np}{(n-kp)},\delta}\leq C\|f\|_{k,q,\delta},\ \ \mbox{if}\ \ n-kp>0\ \ \mbox{and}\ \ 
                                                                                 p\leq q\leq\frac{np}{(n-p)}\\
\mbox{(ii)}&\|f\|_{\infty ,\delta}\leq C\|f\|_{k,p,\delta},\ \ \mbox{if}\ \ n-kp<0,\ \ \mbox{and}\\
\, & |f(x)| = o(r^\delta )\ \ \mbox{as}\ \ r\to\infty.\end{array}\] }

Another tool we will require is the ``multiplication lemma". In what follows we will be mainly using Sobolev spaces with $p=2$, and it is for this case that we give the multiplication lemma (adapted from \cite{CC:sob}).

{\bf Multiplication Lemma}

{\em Pointwise multiplication on $\Rnos^n$ is a continuous bilinear mapping
\[ W^{k_1,2}_{\delta_1}\times W^{k_2,2}_{\delta_2}\to W^{k,2}_\delta ,\]
if $k_1,k_2\geq k,\, \, \, k<k_1+k_2-n/2$, and $\delta >\delta_1+\delta_2$. }

In discussing the decay conditions on $\psi$ it will be useful to have some new notation. Firstly, we introduce two new 2-spinors $o_A$ and $\iota_A
$ by,
\begin{eqnarray}\label{oi-np}
\nonumber &&u_A=\sqrt{R}e^{\frac{1}{2}i\chi}o_A\\
\nonumber &&v_A=\sqrt{R}e^{\frac{1}{2}i\chi}\iota_A ,\\
          &&\mbox{with}\,\,\, \iota^Co_C =1.
\end{eqnarray}
Here, $R$ is a positive real function and $\chi$ a real function (in fact, as it turns out, the same function introduced in the static case). Our 
non-degeneracy condition now reads
\[ v^Au_A =Re^{i\chi}\neq 0.\]
The dyad ${o_A,\iota_A}$ give a spinor dyad basis which is ``co-moving" with the ``Dirac flow" given by $j^\alpha$. In general we have 
\[j^\alpha =R\sigma^\alpha_{A\dot{A}}\left( o^A\bar{o}^{\dot{A}}+\iota^A\bar{\iota}^{\dot{A}}\right),\]
with $j^\alpha j_\alpha =2R^2$. These ideas can be developed further and lead to a Newman-Penrose type formalism for the Maxwell-Dirac system (for the 
Newmann-Penrose formalism in General Relativity see \cite{PR:book}). 
 We will not fully pursue this here, although we go somewhat down this path in our proof of lemma 2, below.

For our static systems we have
\begin{eqnarray}\label{TR-oi}
\nonumber && R=X^2+Y^2\\
\nonumber && \vb{V}=R\uvb{V}=R(\sin\tau\,\cos\eta\, ,\sin\tau\,\sin\eta\, ,\cos\tau)\\
\nonumber && X=\sqrt{R}\cos\frac{\tau}{2}\\
\nonumber && Y=\sqrt{R}\sin\frac{\tau}{2}\\
\nonumber && (o_A) =\left(\begin{array}{c}
                                         \sin\frac{\tau}{2}\, e^{-\frac{i}{2}\eta}\\
                                         -\cos\frac{\tau}{2}\, e^{\frac{i}{2}\eta}
                          \end{array}\right)\\
          &&  (\iota^A) =\left(\begin{array}{c}
                                         \sin\frac{\tau}{2}\, e^{\frac{i}{2}\eta}\\
                                         -\cos\frac{\tau}{2}\, e^{-\frac{i}{2}\eta}
                          \end{array}\right)
\end{eqnarray}
Notice that $o_A$ and $\iota^A$ are both $O(1)$ as $r\to\infty$, so $\psi$ decays as $\sqrt{R}$.

In discussing decay conditions on the $A^\alpha$ we need, of course to be aware that $A^\alpha$ is defined only up to a gauge transformation. This problem 
is usually resolved (to some extent) by imposing gauge conditions, such as the Lorenz gauge. We restrict our attention to stationary gauges and demand that
 $A^0$ be $O(1/r)$ as $r\to\infty$ in some gauge. If we demand that $\vb{A}$ be $O(1/r)$ then the question is, can we solve
\[ \nabla^2\phi +\vb{\nabla .A}=0,\]
so that $\vb{A}'=\vb{A}+\vb{\nabla}\phi$ is $O(1/r)$? That is, can we find a gauge transformation which takes us to the Lorenz gauge with the new $\vb{A}$ 
still satisfying the appropriate $O(1/r)$ decay? The answer is yes provided we choose the original $\vb{A}$ in the correct function space.

We need $A^\alpha$ to be at least twice (weakly) differentiable to make sense of the Maxwell equations. So to get the appropriate differentiability and 
decay we will take (see definition, below)
\[ A^\alpha\in W^{2,p}_{-1+\epsilon}(E_\rho ),\]
for some $p>3/2$ and all $\epsilon >0$. Here, $E_\rho =\Rnos^3\backslash B_\rho$, with $\rho <\infty$ large enough so that $B_\rho$ encloses all external sources. 
This means we can now solve the gauge equation for $\phi\in W^{3,p}_\epsilon (E_\rho ) $, with $\vb{A}'=\vb{A}+\vb{\nabla}\phi\in W^{2,p}_{-1+\epsilon}
(E_\rho )$, in fact, for $0<\epsilon <1$ the Laplacian gives an isomorphism between the function spaces $W^{3,p}_\epsilon$ and $W^{1,p}_{-2+\epsilon}$, see 
\cite{RB:mass}.

The Maxwell equations imply that $j^\alpha \in W^{0,p}_{-3+\epsilon}(E\rho )$. In the vector basis co-moving with $j^\alpha$ (induced from the co-moving 
dyad) the charge density is $\sqrt{2}R$, we would expect therefore that $R$ is $O(1/r^3)$ as $r\to\infty$. We also require at least three derivatives for
 $R$ to define the Maxwell equations when the $A^\alpha$ are written in terms of the components of $\psi$. This suggests we should take 
$R\in W^{3,p}_{-3+\epsilon}(E_\rho )$. The $o_A$ and $\iota_A$ must also have at least three (weak) derivatives and we also need to ensure that 
$j^\alpha\in W^{0,p}_{-3+\epsilon}(E_\rho )$. We will require $o_A ,\iota_A\in W^{3,p}_\epsilon (E_\rho )$, for any $\epsilon >0$. This leaves the 
differentiability and decay of $\chi$ to be determined. Again we will require at least three derivatives of $\chi$. The decay rate, however must be 
determined from the equations. 

We can now give our definition of an isolated Maxwell-Dirac system. For concreteness and ease of manipulation we will restrict our attention to Sobolev spaces 
$W^{k,2}_\delta (E_\rho )$.
\begin{definition} A stationary Maxwell-Dirac system will be said to be isolated if, in some stationary gauge, we have
\[\psi = e^{iEt}R\left(\begin{array}{c}
                                      e^{\frac{i\chi}{2}}o_A\\
                                      e^{-\frac{i\chi}{2}}\bar{\iota}_{\dot{A}}
                       \end{array}\right),\]
with $E$ constant and
\begin{eqnarray*}
&&R\in W^{3,2}_{-3+\epsilon}(E_\rho );\,\,\, o_A,\iota_A\in W^{3,2}_\epsilon (E_\rho )\,\,\, 
\mbox{and}\,\,\, A^\alpha\in W^{2,2}_{-1+\epsilon}(E_\rho ),\,\,\,
\mbox{for some}\,\,\, \rho >0\\
&&\mbox{and any}\,\,\,\epsilon >0.
\end{eqnarray*}
\end{definition}
\underline{Remark}
\newline This definition ensures, after use of the Sobolev inequality and the multiplication lemma, that $\psi =o(r^{-\frac{3}{2}+\epsilon})$ and $A^\alpha =
o(r^{-1+\epsilon})$.

We are now in a position to prove a lemma which will be used in the proof of theorem 1 in the next section. But we first need some new notation.

We introduce the complex null tetrad vectors
\begin{eqnarray*}
l^\alpha =\sigma^\alpha_{A\dot{A}}o^A\bar{o}^{\dot{A}}&,&n^\alpha =\sigma^\alpha_{A\dot{A}}\iota^A\bar{\iota}^{\dot{A}}\\
m^\alpha =\sigma^\alpha_{A\dot{A}}o^A\bar{\iota}^{\dot{A}}&,&\bar{m}^\alpha =\sigma^\alpha_{A\dot{A}}\iota^A\bar{o}^{\dot{A}}.
\end{eqnarray*}
This null tetrad can now be used to define the following (Newman-Penrose) intrinsic derivatives
\begin{eqnarray*}
D =l^{\alpha}\frac{\p}{\p x^\alpha} &,& \bigtriangleup = n^\alpha\frac{\p}{\p x^\alpha}\\
\delta =m^{\alpha}\frac{\p}{\p x^\alpha} &,& \bar{\delta} = \bar{m}^\alpha\frac{\p}{\p x^\alpha}
\end{eqnarray*}
With this notation and the expression for $A^{A\dot{A}}$ of section II we find that the (real) potential $A^\alpha$ has to have the following components with 
respect to the null tetrad
\begin{eqnarray}\label{As-np}
\nonumber A_l&=&l_\alpha A^\alpha =\frac{1}{2e}\left[\sqrt{2}m\cos\chi\, -\bigtriangleup\chi +i(\mu-\bar{\mu}+\gamma -\bar{\gamma})\right]\\
\nonumber A_n&=&n_\alpha A^\alpha =\frac{1}{2e}\left[\sqrt{2}m\cos\chi\, +D\chi +i(\rho-\bar{\rho}+\bar{\varepsilon} -\varepsilon )\right]\\
\nonumber A_{m}&=&m_\alpha A^{\alpha}=\frac{i}{2e}\left[ -\frac{\delta R}{R}+\bar{\alpha}-\beta +\tau -\bar{\pi}\right]\\
          A_{\bar{m}}&=&\bar{m}_\alpha A^{\alpha}=\frac{i}{2e}\left[\frac{\bar{\delta} R}{R}-\alpha +\bar{\beta}-\bar{\tau}+\pi\right]
\end{eqnarray}
Here $\alpha ,\beta , \tau ,\mu ,\rho ,\gamma$ and $\varepsilon$ are the NP spin coefficients (Ricci rotation coefficients for the non-holonomic NP 
tetrad), see \cite{PR:book}. Their 
exact form is not important here, what we do need to know is that they are all of the form $o\p o,\iota\p\iota ,
o\p\iota$ or $\iota\p o$, where $\p$ is any one of the NP intrinsic derivatives, $D,\bigtriangleup ,\delta$ or $\bar{\delta}$. Notice that here we are using a 
gauge which has the factor $e^{iEt}$ of definition 4 removed -- for the stationary case this means all variables are independent of $t$ and $A^0\to E/e$ as $r\to \infty$.

\begin{lemma}
For an isolated, stationary Maxwell-Dirac system the following must hold
\[\left.\begin{array}{c}
                               \delta R/R-\frac{E}{e}m_0\\
                                \bar{\delta} R/R-\frac{E}{e}\bar{m}_0
  \end{array}\right\}\in W^{2,2}_{-1+\epsilon}(E_\rho ),\]
for any $\epsilon >0$ and some constant $E$.
\end{lemma}
\begin{proof}
A straightforward application of the multiplication lemma.
\newline With $o_A ,\iota_A\in W^{3,2}_\epsilon (E_\rho )$ and $A^0-E/e, \vb{A}\in W^{2,2}_{-1+\epsilon}(E_\rho )$, for any $\epsilon >0$, we have 
\[A_m-m_0\frac{E}{e} ,A_{\bar{m}}-\bar{m}_0\frac{E}{e}\in W^{2,2}_{\delta_1}(E_\rho ),\,\,\,\delta_1 >-1+3\epsilon\]
and
\[\alpha ,\beta ,\gamma ,\tau ,\mu ,\pi ,\varepsilon\in W^{2,2}_{\delta_2}(E_\rho ),\,\,\,\delta_2 >-1+4\epsilon .\]
The result then follows from \eqref{As-np}.
\end{proof}

{\bf IV. Vanishing Total Charge}

We will be working with the stationary, static Maxwell-Dirac equations. From section II they are
\begin{eqnarray}\label{st-stat}
\nonumber &&(\vb{\nabla}\chi)\vb{\times V}=\vb{0}\\
\nonumber &&\vb{\nabla .V}=-2m(X^2+Y^2)\sin\chi\\
\nonumber && A^0=\frac{m}{e}\cos\chi +\frac{(\vb{\nabla}\chi)\vb{.V}}{2e(X^2+Y^2)}\\
\nonumber && \vb{A}=\frac{1}{2e(X^2+Y^2)}\left[ (X^2-Y^2)\vb{\nabla}\eta-\vb{\nabla\times V}\right]\\
\nonumber &&\nabla^2 A^0=4\pi ej^0\\
          &&\vb{\nabla\times}(\vb{\nabla\times A})=0.
\end{eqnarray}
We can now state and prove our theorem of vanishing total charge.
\begin{theorem}
An isolated, stationary, static Maxwell-Dirac system is electrically neutral.
\end{theorem}
\begin{proof} A more restricted version of this theorem was proved in \cite{HB:thesis} by one of us (H.B.). 

The stationary gauge of definition 4 is the one for which $A^0\to 0$ as $r\to 0$, the stationary gauge used in equations \eqref{st-stat} has $\psi$ 
independent of $t$. A gauge transformation of the type $\psi\to e^{-iEt}\psi$ will bring the gauge of definition 4 into that of equations 
\eqref{st-stat}. The $A^0$ of these equations then differs by a constant, $-E/e$, from the $A^0$ of definition 4. In fact, we will be able to determine
 $E$ in the proof of the theorem, see also corollary 1.
As we noted earlier we have
\begin{eqnarray*}
o_A&=&\left(\begin{array}{c}\sin\frac{\tau}{2}\, e^{-i\frac{\eta}{2}}\\
                            -\cos\frac{\tau}{2}\, e^{i\frac{\eta}{2}}
            \end{array}\right)\\
\iota^A&=&\left(\begin{array}{c}\sin\frac{\tau}{2}\, e^{i\frac{\eta}{2}}\\
                            -\cos\frac{\tau}{2}\, e^{-i\frac{\eta}{2}}
            \end{array}\right)\\
\end{eqnarray*}
in the static case. The system is isolated so using the multiplication lemma we have that
\[ \sin\tau\, ,\cos\tau\, ,\sin\eta\, ,\cos\eta\, \in W^{3,2}_\epsilon (E_\rho ).\]
Now we have $m_0=\sigma_0^{A\dot{A}}o_A\bar{\iota}_{\dot{A}}=0$, so using lemma 2 we have
\[ \frac{\delta R}{R}\,\,\,\mbox{and}\,\,\,\frac{\bar{\delta}R}{R}\in W^{2,2}_{-1+\epsilon}(E_\rho ).\]
Which in our static case gives,
\begin{eqnarray}\label{R-dec}
\nonumber &&\cos\tau\,\left(\cos\eta\,\frac{\p R}{\p x}+\sin\eta\,\frac{\p R}{\p y}\right) -\sin\tau\,\frac{\p R}{\p z}\in W^{2,2}_{-1+\epsilon}(E_\rho )\\
\mbox{and}\,\,\,&&\sin\eta\,\frac{\p R}{\p x}-\cos\eta\,\frac{\p R}{\p y}\in W^{2,2}_{-1+\epsilon}(E_\rho ).
\end{eqnarray}
The second of equations \eqref{st-stat} can be written as
\begin{eqnarray}\label{sn-eq}
 -2m\sin\chi\, &=&\frac{\vb{\nabla .V}}{R}=\frac{\uvb{V}\vb{.\nabla}R}{R}+\vb{\nabla .}\uvb{V},
\end{eqnarray}
with $\uvb{V}=(\sin\tau\,\cos\eta\, ,\sin\eta\sin\tau\, ,\cos\tau )$. Using the multiplication lemma we have
\[\uvb{V}\in W^{3,2}_{2\epsilon}(E_\rho )\,\,\,\mbox{and}\,\,\,\vb{\nabla .}\uvb{V}\in W^{2,2}_{-1+2\epsilon}(E_\rho ),\]
for any $\epsilon >0$. We also have
\begin{eqnarray}\label{dd-R}
\frac{\uvb{V}\vb{.\nabla}R}{R}=\sin\tau\,\cos\eta\,\frac{\p R}{\p x}+\sin\tau\,\sin\eta\,\frac{\p R}{\p y}+\cos\tau\,\frac{\p R}{\p z}.
\end{eqnarray}
Next we utilise the invariance of our equations under Lorentz transformations. In fact, we know that if $(u_A(x^\alpha ),v_A(x^\alpha ) )$ is a
solution to the Maxwell-Dirac equations then $(u_A(\h{x}^\alpha ),v_A(\h{x}^\alpha ))$ is a solution to the original system in the $x^\alpha$ coordinates; here $\h{x}^\alpha$ are 
the Lorentz transformed Cartesian coordinates. This is, of course, true for any linear Lorentz invariant theory.
Consider the rotation
\begin{eqnarray*}
&&\h{x}=x\cos\omega\, +y\sin\omega\\
&&\h{y} =-x\sin\omega\, +y\cos\omega\\
&&\h{z}=z,\,\,\,\mbox{with}\,\,\, \omega =-\frac{\pi}{2}.
\end{eqnarray*}
This gives
\[\frac{\sin\eta (\h{x}^\alpha )}{R(\h{x}^\alpha )}\,\frac{\p R(\h{x}^\alpha )}{\p x}-\frac{\cos\eta (\h{x}^\alpha )}{R(\h{x}^\alpha )}\,\frac{\p R(\h{x}^\alpha )}{\p y}=
  \frac{\cos\eta (\h{x}^\alpha )}{R(\h{x}^\alpha )}\,\frac{\p R(\h{x}^\alpha )}{\p \h{x}}+\frac{\sin\eta (\h{x}^\alpha )}{R(\h{x}^\alpha )}\,\frac{\p R(\h{x}^\alpha )}{\p \h{y}}\]
Diffeomorphisms $\Rnos^3\to \Rnos^3$ induce an isomorphism of the Sobolev spaces $W^{k,p}_\delta$, see \cite{CC:sob}. The rotation above preserves $E_\rho $ and will 
give an isomorphism of the Sobolev spaces $W^{k,p}_{\delta}(E_\rho )$. Using our last equation and \eqref{R-dec} gives
\[\frac{\cos\eta}{R}\,\frac{\p R}{\p x}+\frac{\sin\eta}{R}\,\frac{\p R}{\p y}\in W^{2,2}_{-1+\epsilon}(E_\rho ).\]
This equation and \eqref{R-dec} with the multiplication lemma then gives (multiply each equation in turn by $\sin\eta$ and $\cos\eta$ etc.),
\[\frac{1}{R}\frac{\p R}{\p x},\frac{1}{R}\frac{\p R}{\p y}\in W^{2,2}_{-1+\epsilon}(E_\rho ).\]
The rotation
\[ \h{x} =-z,\,\,\, \h{y}=y\,\,\, \h{z}=x\]
gives
\[\frac{\sin\eta (\h{x}^\alpha )}{R(\h{x}^\alpha )}\,\frac{\p R(\h{x}^\alpha )}{\p x}-\frac{\cos\eta (\h{x}^\alpha )}{R(\h{x}^\alpha )}\,\frac{\p R(\h{x}^\alpha )}{\p y}=
  \frac{\sin\eta (\h{x}^\alpha )}{R(\h{x}^\alpha )}\,\frac{\p R(\h{x}^\alpha )}{\p \h{z}}-\frac{\cos\eta (\h{x}^\alpha )}{R(\h{x}^\alpha )}\,\frac{\p R(\h{x}^\alpha )}{\p \h{y}}.\]
Again using the multiplication lemma with \eqref{R-dec}, we have
\[\frac{\sin\eta}{R}\,\frac{\p R}{\p z}\in W^{2,2}_{-1+\epsilon}(E_\rho ).\]
In the same fashion, using the rotation $\h{x}=x,\,\,\, \h{y}=z,\,\,\, \h{z}=-y$, we have
\[\frac{\cos\eta}{R}\,\frac{\p R}{\p z}\in W^{2,2}_{-1+\epsilon}(E_\rho ).\]
Another use of the multiplication lemma with the last two equations gives $\frac{1}{R}\p R/\p z\in W^{2,2}_{-1+\epsilon}(E_\rho )$.

Altogether we have
\[ \frac{1}{R}\frac{\p R}{\p x},\,\,\,\frac{1}{R}\frac{\p R}{\p y},\,\,\, \frac{1}{R}\frac{\p R}{\p z}\in W^{2,2}_{-1+\epsilon}(E_\rho )\]
A final use of the multiplication lemma with \eqref{R-dec} and we have
\[\frac{\uvb{V}\vb{.\nabla}R}{R}\in W^{2,2}_{-1+\epsilon}(E_\rho ).\]
We can now conclude from \eqref{sn-eq} that $\sin\chi\,\in W^{2,2}_{-1+\epsilon}(E_\rho )$, for any $\epsilon >0$. By the Sobolev inequality $\sin\chi\, =o(r^{-1+\epsilon})$ as 
$r\to\infty$.

Now sine is an invertible $C^\infty$ function on the range of $\chi$ (on $E_\rho$, with $\rho$ large) with $\sin\chi =0$ for $\chi =0$. So we can now write
\begin{eqnarray}\label{ab-chi}
\nonumber &&\chi = n\pi +\mu,\ \ \mbox{with}\ \ n=0,\pm 1,\pm 2,\ldots\ \ \mbox{and}\\
 &&\mu\in W^{2,2}_{-1+\epsilon}(E_{\rho}).
\end{eqnarray}
Next we use the first of equations \eqref{st-stat} to rewrite $A^0$ entirely in terms of $\chi$ and $|\nabla\chi|$.
 This equation implies we may write $\vb{V}=\gamma\vb{\nabla}\chi$, for some function 
$\gamma$. We also have $|\vb{V}|=X^2+Y^2=|\gamma|\,|\nabla\chi|$, so that
\[\nonumber A^0=\frac{m}{e}\cos\chi\, +\frac{\varepsilon}{2e}|\nabla\chi |,\ \ \mbox{with}\ \ 
\varepsilon =\frac{\gamma}{|\gamma|}.\]
Which, using \eqref{ab-chi}, may be written as
\[ A^0-\frac{m}{e}\cos (n\pi )\, =-\frac{2m}{e}\cos (n\pi )\,\sin^2(\frac{\mu}{2})\, +
\frac{\varepsilon}{2e}|\nabla\mu |.\]
Hence,
\[ A^0-\frac{m}{e}\cos (n\pi )\,\in W^{1,2}_{-2+\epsilon}(E_{\rho}).\]
Note that $|\nabla\chi |$ is bounded on $E_\rho$. So $\gamma$ cannot change sign on $E_\rho$ as $R=X^2+Y^2\neq 0$, $\varepsilon$ is fixed.

From the (first) Sobolev inequality applied to $A^0-\frac{m}{e}\cos n\pi$, with $p=q=2$, we have that
\[ A^0-\frac{m}{e}\cos (n\pi )\,\in W^{0,6}_{-2+\epsilon}(E_{\rho}).\]
Consequently, we also have
\[ A^0-\frac{m}{e}\cos (n\pi )\,\in W^{0,6}_{-\frac{5}{4}}(E_{\rho}).\]
Now, in the static case, $j^0=\sqrt{2}R$, so $\nabla^2(A^0-\cos n\pi\, )\in W^{3,2}_{-3+\epsilon}$. Hence,
\[ A^0-\frac{m}{e}\cos (n\pi )\,\in W^{5,2}_{-1+\epsilon}(E_{\rho})\]
and from the Sobolev inequality we find that $|\p_i A^0| <Cr^{-2+\epsilon}$ (with $\p_i=\p /\p x^i$). So we have,
\[ |\p_iA^0|^6r^{-6(-1-\frac{5}{4})-3}<C^4|\p_iA^0|^2r^{-2(-3+\epsilon)-3},\]
for any $0<\epsilon <1/12$. Thus, $\p_iA^0\in W^{0,6}_{-9/4}(E_\rho )$. Which finally gives us,
\[ A^0-\frac{m}{e}\cos (n\pi )\,\in W^{1,6}_{-\frac{5}{4}}(E_{\rho}),\]
and the Sobolev inequality now gives
\[ A^0-\cos n\pi\, =o(r^{-\frac{5}{4}}).\]
From which it is clear that the total electric charge of the system
\[ \lim_{\rho\to\infty}\frac{1}{4\pi}\int_{S_{\rho}}\left(\vb{\nabla}A^0\right)\vb{.}d\vb{S},\ \ 
\mbox{with}\ \ S_{\rho}\ \ \mbox{the sphere of radius $\rho$},\]
must vanish.
\end{proof}
\begin{corollary}
In the gauge in which $A^0\to 0$ as $r\to\infty$ the Dirac bi-spinor of an isolated, stationary, static Maxwell-Dirac
 system takes the form
\[ \psi =e^{\pm imt}\phi,\ \ \mbox{with}\ \ \phi\in W^{3,2}_{-\frac{3}{2}+\epsilon}(E_\rho)\ \ \mbox{(as above)}.\]
\end{corollary}
\begin{proof}
This result is a simple consequence of the proof of the theorem. We had
\[ A^0 -\frac{m}{e}\cos (n\pi )\,\in W^{1,2}_{-2+\epsilon}(E_\rho ) \ .\]
The constant term, $-\frac{m}{e}\cos n\pi\, $ is removed by the gauge transformation
\[ \psi\to e^{-i\cos (n\pi )\, mt}\psi .\]
\end{proof}
In \cite{CR:pap1} we presented a unique spherically symmetric solution which provides a good example of the properties just described. 
For large $r$ we found
\begin{eqnarray*}
&&\chi =\pi -\frac{1}{mr}-\frac{1}{168m^3r^3}+O(\frac{1}{r^6})\\
&& A^0 =-\frac{m}{e}+\frac{1}{emr^2}-\frac{3}{112em^3r^4}+O(\frac{1}{r^6}).
\end{eqnarray*}

{\bf V. The Axially Symmetric Case}

By axially symmetric we mean that the system is invariant under rotations about a fixed axis. This requires that gauge invariant quantities be invariant 
under translations in the azimuthal coordinate $\phi$.
\begin{definition}
A Maxwell-Dirac system will be called axially symmetric if
\begin{eqnarray*}
&&\left[ L,\frac{\p}{\p \phi}\right] =\left[ N,\frac{\p}{\p \phi}\right]=0,\\ \mbox{where} && 
 L =\sigma^{\alpha}_{A\dot{A}}u^A\bar{u}^{\dot{A}}\frac{\p}{\p x^{\alpha}}\ \ \mbox{and}\\
&& N  =\sigma^{\alpha}_{A\dot{A}}v^A\bar{v}^{\dot{A}}\frac{\p}{\p x^{\alpha}}.
\end{eqnarray*}
\end{definition}
For our static systems we require only that $L$ be invariant under translations in $\phi$. In fact, writing $L$ in cylindrical polar coordinates,
\begin{eqnarray*}
L&=&L^0\frac{\p}{\p t}+\frac{1}{\sqrt{2}}\left( V^{\rho}\uvb{\rho}+V^{\phi}\uvb{\phi}+V^{z}\uvb{k}\right)\\
    &=&L^0\frac{\p}{\p t}+\frac{1}{\sqrt{2}}\left( V^{\rho}\frac{\p}{\p\rho}+V^{\phi}\frac{\p}{\p\phi}+V^{z}\frac{\p}{\p z}\right),
\end{eqnarray*}
we find that
\begin{eqnarray*}
&&L^0 =\frac{X^2+Y^2}{\sqrt{2}},\ \ V^{\rho} =2XY\cos(\eta -\phi ),\\
&& V^{\phi} =2XY\sin (\eta -\phi )\ \ \mbox{and}\ \ V^z =X^2-Y^2
\end{eqnarray*}
must all be independent of $\phi$. This means our static Maxwell-Dirac system is axially symmetric if
\[ \frac{\p X}{\p\phi}=\frac{\p Y}{\p\phi}=\frac{\p (\eta -\phi )}{\p\phi}=0.\]
This information lets us characterise stationary, axially symmetric, static Maxwell-Dirac systems as follows.
\begin{lemma}
A non-trivial static, axially symmetric Maxwell-Dirac system is stationary if and only if $\eta =\phi$ in the gauge given in \eqref{f-gauge}.
\end{lemma}
\begin{proof}
As $\vb{V}$ is independent of $\phi$ then so is $\chi$, as $\sin\chi\, =(\vb{\nabla .V})/(X^2+Y^2)$. The reality 
condition,
\[ \frac{\p\vb{V}}{\p t}+(\vb{\nabla}\chi)\vb{\times V}=\vb{0}\]
gives,
\begin{eqnarray*}
\frac{\p V^{\rho}}{\p t}-V^{\phi}\frac{\p\chi}{\p z}&=&0\\
\frac{\p V^{\phi}}{\p t}-V^{z}\frac{\p\chi}{\p\rho}+V^{\rho}\frac{\p\chi}{\p z}&=&0\\
\frac{\p V^{z}}{\p t}+V^{\phi}\frac{\p\chi}{\p\rho}&=&0.
\end{eqnarray*}
If the system is stationary lemma 1 says $\vb{V}$ is independent of $t$. So either $V^{\phi}=0$ or $\chi$ is constant. Constant $\chi$ leads 
to the trivial solution with $X=Y=0$. So we must take $V^{\phi}=2XY\sin (\eta -\phi)\, =0$. We have $\eta =\phi$ mod $n\pi$.

On the other hand,
 if $\eta =\phi$ we have $V^{\phi}=0$ and consequently both $V^{\rho}$ and $V^z$ are independent of $t$. Combining these with $\p_t (X^2+Y^2)=0$
 gives the result.
\end{proof}
From now on we study the axially symmetric, stationary, static Maxwell-Dirac equations.

It will prove convenient in what follows to use spherical polar coordinates $(r,\theta ,\phi )$ and to make the following 
change of variables
\[ X=\sqrt{R}\cos(\frac{\tau}{2})\ \ \mbox{and}\ \ Y=\sqrt{R}\sin(\frac{\tau}{2}).\]
All our dependent variables depend only on $(r,\theta)$, the equations are
\begin{eqnarray}\label{axist-stat}
\nonumber&&\vb{V}=R\left[\cos(\tau -\theta)\, \uvb{r}+\sin(\tau -\theta)\,\uvb{\theta}\right]\\
\nonumber&& (\vb{\nabla}\chi)\vb{\times V}=\vb{0}\\
\nonumber&& \vb{\nabla .V}=-2mR\sin\chi\\
\nonumber&& A^0=\frac{m}{e}\cos\chi\, +\frac{\varepsilon}{2e}|\nabla\chi |\\
         && \vb{A}=\frac{1}{2e}\left\{ \frac{\cos\tau}{r\sin\theta}
-\frac{1}{rR}\left[\frac{\p}{\p r}
\left( rR\sin(\tau -\theta)\right)-\frac{\p}{\p\theta}\left( R\cos(\tau -\theta)\right)\right]\right\}\uvb{\phi},
\end{eqnarray}
together with the Maxwell equations. Here $\uvb{r}$, $\uvb{\theta}$ and $\uvb{\phi}$ are the unit coordinate 
vectors. 

We note that $A^\alpha$ automatically satisfies the Lorenz gauge condition.

In the spherically symmetric case \cite{CR:pap1} we have $\tau =\theta$ with $R$ and $\chi$ functions of $r$
 only. In which case 
\[ \vb{A}=\frac{1}{2e}\frac{\cot\theta}{r}\uvb{\phi},\]
the magnetic monopole.

Other tractable cases are those for which $\tau$ is constant. It is straightforward to show there are really 
only two cases, see \cite{HB:thesis},
\begin{itemize}
\item the cylindrically symmetric case, $\tau=\pi /2$, see \cite{HC:pap2}.
\item the case $\tau =0$, variables depend on $z$ only, see \cite{CR:pap4}.
\end{itemize}

{\bf VI. Magnetic Monopoles}

The spherically symmetric solution has an external (i.e. not sourced by the Dirac field directly) 
electrically charged magnetic monopole. In this section we will prove a theorem which shows that this is, 
to some extent, the generic situation. We will show that the axially symmetric, stationary, static Maxwell-Dirac 
system can have an external Coulomb point charge only if it is magnetically charged. First we define what 
we mean by an external Coulomb field.
\begin{definition}
We will say that a Maxwell-Dirac system has an external Coulomb field if we can choose spherical polar coordinates
 and a ball $B_\rho$ centred at $r=0$ such that
\[ A^0 =\frac{q}{r}+h,\ \ \mbox{in}\ \ B_\rho ,\ \ \rho >0,\]
with, $h$, a bounded function on $B_\rho$ and $q$ constant.
\end{definition}

\underline{Remarks}
\begin{enumerate}
\item $q/r$ is of course harmonic on $B_\rho\backslash\{0\}$, so it is not directly sourced by the Dirac field
 via the Maxwell equation $\nabla^2A^0 =4\pi ej^0$. In this sense the Coulomb field is ``external" to the 
Dirac field.
\item The condition that $h$ is bounded on $B_\rho$ is quite weak. In practice it will follow from elliptic 
regularity of the Poisson equation $\nabla^2h=4\pi e j^0$ ($j^0 =\sqrt{2}(X^2+Y^2)=\sqrt{2}R$ must be at least
$L^1(B_\rho)$ for the total Dirac charge to be well-defined on $B_\rho$), see for example \cite{GT:book} 
and \cite{LL:book}. We also require $R$ to be differentiable at least three times (if only in the weak sense) to 
satisfy the Maxwell equation for $\vb{A}$  -- this puts $R$ in $W^{3,p}_\delta$ for some $p\geq1$. 
Consequently, $h$ will be in 
$W^{5,p}_{\delta +2}$, which ensures that $h$ can be included in one of the classical weighted function spaces. 
\end{enumerate}
Now for our theorem.
\begin{theorem}
Suppose an axially symmetric, stationary, static Maxwell-Dirac system has an external Coulomb field. Let 
$A_{(\rho ,\rho_1)} =B_\rho\backslash B_{\rho_1}$, with $\rho >\rho_1>0$ and $B_\rho$ as in the definition 
above. If 
$r\p_r\tau -\p_{\theta}R/R$ is bounded on $A_{(\rho ,\rho_1)}$, then the Coulomb point charge necessarily 
carries a magnetic charge of 
Dirac value, $\frac{\pm 1}{2e}$; i.e. all electric point charges also carry a magnetic monopole.
\end{theorem}

\underline{Remark}

The condition $r\p_r\tau -\p_{\theta}R/R$ is bounded on $A_{(\rho ,\rho_1)}$ is not particularly strong. It 
is true, for example, if $X,Y\in C^1_0(A_{(\rho ,\rho_1)})$ -- we assume $R=X^2+Y^2\neq 0$; or, when $\vb{A}\in W^{2,2}_{1}(B_\rho\backslash\{0\})$.

\begin{proof}
In \cite{HB:thesis} it was shown that if a Maxwell-Dirac system has a central Coulomb charge then the magnetic field is necessarily unbounded at the 
centre.

In our proof of theorem 1 we noted that $\vb{V}=\gamma \vb{\nabla}\chi$, so from \eqref{axist-stat} we have
\begin{eqnarray}\label{cos-sin}
\nonumber \cos (\tau -\theta )\, &=&\varepsilon\frac{\p_r\chi}{|\nabla\chi |}\ \ \mbox{and}\\
          \sin (\tau -\theta )\, &=&\varepsilon\frac{\p_\theta\chi}{r|\nabla\chi |},\ \ \mbox{where}\ \ 
\varepsilon =\frac{\gamma}{|\gamma |}.
\end{eqnarray}

We will first show that $\sin(\tau -\theta )\to 0$ as $r\to 0$.

Now, $A^0=\frac{q}{r}+h$, so from \eqref{axist-stat} we have 
\begin{equation}\label{nab-chi} |\nabla\chi|=\frac{q}{r}+g,\end{equation}
where $g=h-\frac{m}{e}\cos\chi$ is bounded on $B_\rho$. 
Next, we write 
\[ \chi =q\ln r\, +\zeta ,\ \ \mbox{on}\ \ B_\rho\backslash \{0\}.\]
Equation \eqref{nab-chi} is
\begin{equation}\label{zeta-eq}(\p_r\zeta)^2+\frac{2q}{r}\p_r\zeta +\frac{1}{r^2}(\p_\theta\zeta)^2 =
\frac{2q}{r}g+g^2 .\end{equation}
Note that $r|\nabla\chi |\to q$ as $r\to 0$ so that $r\p_r\chi = q+r\p_r\zeta$ and $\p_\theta\chi =\p_\theta\zeta$
are bounded on $A_{(\rho ,\rho_1)}$, for $\rho$ small enough. Let $\eta$ be the smallest non-negative number 
such that $r^\eta\p_r\zeta\to 0$ as $r\to 0$. We can write \eqref{zeta-eq} as
\[ (r^{\frac{1+\eta}{2}}\p_r\zeta)^2+2qr^\eta\p_r\zeta+(r^{\frac{\eta -1}{2}}\p_\theta\zeta)^2 =
2qr^\eta g+r^{\eta +1}g^2.\]
It is clear that $r^{\frac{1+\eta}{2}}\p_r\zeta\to 0$ and $r^{\frac{\eta -1}{2}}\p_\theta\zeta\to 0$. From which
 we conclude that $\eta\leq 1$ and so $\p_\theta\chi =\p_\theta\zeta\to 0$ as $r\to 0$. Consequently,
\[ \sin(\tau -\theta )\, =\frac{\p_\theta\chi}{r|\nabla\chi |}\to 0,\ \ \mbox{as}\ \ r\to 0.\]
Now $r\p_r\chi\to q$ as $r\to 0$, so $r\p_r\chi$ cannot change sign as $\theta$ varies, for $\rho$ small.
Writing $\varepsilon_1 = q/|q|$ we have, from \eqref{cos-sin},
\begin{equation}\label{cos-s}\left.\begin{array}{c}
\cos\tau (r,\pi )\to -\varepsilon\varepsilon_1\\
      \cos\tau (r,0)\to \varepsilon\varepsilon_1\end{array}\right\} \ \ \mbox{as}\ \ r\to 0. \end{equation}

We now work on $A_{(\rho ,\rho_1)}$ with $\rho$ small enough that $\cos (\tau -\theta )\neq 0$. The magnetic 
charge of the magnetic field $\vb{B}=\vb{\nabla\times A}$ in $B_r$, $\rho >r >\rho_1$, is
\begin{equation}\label{mag-ch} b=\frac{1}{4\pi}\int_{S_r}\vb{B.}d\vb{S}=\frac{1}{2}\int^{\pi}_{\theta =0}
\p_{\theta}(r\sin\theta\, A)d\theta , \end{equation}
where $\vb{A}=A\uvb{\phi}$ is given in \eqref{axist-stat}. After some manipulation the third equation of 
\eqref{axist-stat} can be written as
\[ \p_\theta\tau -2+\frac{r\p_r R}{R} = \tan(\tau -\theta )\, \left( r\p_r\tau -\frac{\p_\theta R}{R}\right)
-2mr\sin\chi\, -3\cos (\tau -\theta ).\]
So that if $r\p_r\tau -\frac{\p_\theta R}{R}$ is bounded on $A_{(\rho ,\rho_1)}$ then so is 
$\p_\theta\tau -2+\frac{r\p_r R}{R}$. From \eqref{axist-stat} we have
\begin{eqnarray*}
&&r\sin\theta\, A=\frac{1}{2e}\left\{\cos\tau - P\sin\theta\,\right\},\\
&&\mbox{where}\ \ P=\left( r\p_r\tau -\frac{\p_\theta R}{R}\right)\cos (\tau -\theta )\, -
\left(\p_\theta\tau -2 +\frac{r\p_r R}{R}\right)\sin (\tau -\theta ).\end{eqnarray*}
Clearly, under the conditions of the theorem, $P$ is bounded on $A_{(\rho ,\rho_1 )}$. From \eqref{mag-ch} we now have
\begin{eqnarray*}
 b&=&\frac{1}{2e}\left[\frac{\cos\tau(r,\pi )\, -\cos\tau(r,0)}{2}\right]-\left[P\sin\theta\,\right]^{\pi}_{\theta =0}\\
  &=&\frac{1}{2e}\left[\frac{\cos\tau(r,\pi )\, -\cos\tau(r,0)}{2}\right].
\end{eqnarray*}
Finally, from equations \eqref{cos-s} and \eqref{mag-ch} we obtain the magnetic charge in the limit $r\to 0$
\[ b_0 =\frac{-\varepsilon\varepsilon_1}{2e}=\frac{\pm 1}{2e}.\]
\end{proof}
\begin{corollary}
Suppose we have an axially symmetric, isolated, stationary, static Maxwell-Dirac system with the only external sources being $N$ isolated 
electrically charged magnetic monopoles. Let the conditions of theorem 2 apply in the $N$ balls $B_{\rho_i}$, containing the charges. 
Let the conditions of theorem 1 
apply on $E_{\bar{\rho}}$. Then, if $N$ is even there 
are $N/2$ positive charges and $N/2$ negative charges (with corresponding monopoles), with the total magnetic charge of the system being 
zero. If $N$ is odd there are $(N-1)/2$ charges with one 
sign and $(N+1)/2$ charges with the opposite sign, with the total magnetic charge of the system being $\pm 1/(2e)$.
\end{corollary}
\begin{proof}
We have $N$ charged monopoles each in a ball $B_{\rho_i}$, $i=1,2,3\ \ldots\ N$, there are no other external sources and all the $B_{\rho_i}$ 
are properly contained in $B_{\bar{\rho}}$.

We have $\vb{A}$ is $O(1/r)$ so that $P$, as defined in the above proof, is bounded on $E_{\bar{\rho}})$.
Using the results in the proof of theorem 2 and the divergence theorem we find the total magnetic charge of the system.
\begin{eqnarray*}
b_{\mbox{total}}=\lim_{r\to\infty}\frac{1}{4\pi}\int_{S_r}\left(\vb{\nabla\times A}\right)\vb{.}d\vb{S}&=&
\frac{1}{2e}\lim_{r\to\infty}\left[\frac{\cos\tau(r,\pi)-\cos\tau(r,0)}{2}\right]\\
                                                                                 &=& -\frac{\varepsilon}{2e}\sum^N_{i=1}\varepsilon_i.
\end{eqnarray*}
This gives,
\[ 1\geq\varepsilon\sum^N_{i=1}\varepsilon_i\geq -1,\]
from which the results of the corollary follow.
\end{proof}

{\bf VII. Conclusions}

In classical physics one expects stationary or static systems to be the end point of
 some time evolution. This clearly cannot be the case for a single isolated electron modelled by the Maxwell-Dirac system. To 
construct such a model we will have to abandon one, or both, of the static and stationary assumptions. The electron must be a dynamic object. 
The same also applies to other stable objects such as the Hydrogen atom. A Maxwell-Dirac model of such objects must be non-static 
or non-stationary, or both.

Theorem 1 is remarkable in many ways. No matter what arrangement of external electric and magnetic fields inside the ball $B_\rho$, no matter 
what we do to the topology in $B_\rho$ the total electric charge of the system must be zero. The total charge vanishes purely as a 
result of the asymptotic decay and regularity conditions.

\end{document}